\documentclass[11pt,aps,amsmath,amssymb,nofootinbib,notitlepage,longbibliography]{revtex4-1}
\usepackage{amsmath,amssymb}
\usepackage{slashed}
\usepackage{graphicx}
\usepackage{bm}
\usepackage[top=1.0in,bottom=1.0in,left=1.0in,right=1.0in]{geometry}
\usepackage[colorlinks,linkcolor=blue,citecolor=blue]{hyperref}
\usepackage{subcaption}
\newcommand{\be}{\begin{equation}}
\newcommand{\ee}{\end{equation}}
\newcommand{\bea}{\begin{eqnarray}}
\newcommand{\eea}{\end{eqnarray}}
\newcommand{\ba}{\begin{eqnarray}}
\newcommand{\ea}{\end{eqnarray}}

\newcommand{\la}{\label}

\begin{document}

\title{The Nambu-Goto string in QCD:\\
Dipole interactions, scattering and entanglement}


\author{Yizhuang Liu}
\email{yizhuang.liu@uj.edu.pl}
\affiliation{Institute of Theoretical Physics, Jagiellonian University, 30-348 Kraków, Poland}

\author{Maciej A. Nowak}
\email{maciej.a.nowak@uj.edu.pl}
\affiliation{Institute
of Theoretical Physics and Mark Kac Center for Complex Systems Research,
Jagiellonian University, 30-348 Kraków, Poland}

\author{Ismail Zahed}
\email{ismail.zahed@stonybrook.edu}
\affiliation{Center for Nuclear Theory, Department of Physics and Astronomy, Stony Brook University, Stony Brook, New York 11794--3800, USA}



\begin{abstract}
We revisit some aspects of the stringy approach to dipole-dipole interactions, scattering and entanglement
in QCD, using the Nambu-Goto (NG) string,
without recourse to holography.
We first show that the potential between two static dipoles exchanging closed NG strings is attractive at all separations.
Underlining the exchange there is an emergent
entropy, that is dominated by the tachyon at large separation,
and vanishes at short separation, a measure of the confinement-deconfinement transition.  The same tachyon is dominant in the
scattering amplitude, as a correlator of two Wilson loops for two fixed dipole-like hadrons separated at large rapidity gap,
where the contribution of the worldsheet fermions is included.
While the tachyon causes the mean string bit density to grow exponentially with the rapidity,
the  total scattering cross section still satisfies the Froissart bound by quantum shadowing.
The stringy scattering exchange also carries  an entanglement entropy,  that  saturates when the bound is reached.
For hadrons with varying dipole sizes, the tachyon exchange takes place in  hyperbolic space in the conformal
limit. The result  for the full S-matrix is reminiscent of the one from Mueller$^\prime$s evolved dipole wavefunction,
for the total dipole-dipole cross section in perturbative QCD.
\end{abstract}

\maketitle

\section{Introduction}
String theory has been  hailed as a possible consistent theory of quantum gravity. By pertinent compactifications,
it has the potential to lead to  various extensions of the standard model at higher energies. But
perhaps the most compelling application of string theory  may still be in strong interactions, where it originated
from. It is plausible that QCD in the large number of colors $N_c$  limit, may be dual to
an effective string theory with a weak string coupling $g_s\sim \frac 1{N_c}$, as suggested by holography
in higher dimensions~\cite{Ammon:2015wua} (and references therein).

In so far, the compelling arguments for a QCD string stem from lattice QCD simulations~\cite{Kuti:2005xg} (and references therein).
Indeed, detailed QCD  lattice simulations in 1+2 dimensions have shown that the closed flux tube in gauge theories
with various $N_c$, are well described by a Nambu-Goto (NG) string in flat space dimensions~\cite{Athenodorou:2007du}.
Detailed studies of the string mass on its length for various $N_c$, yield results that are in good agreement with the NG string even for short lengths,
well beyond the contribution of the Luscher term.
Although fuzzy, the string is well approximated by a fundamental NG string. The extension of the lattice analysis to 1+3-dimensions for $N_c=3,5$
for the closed string spectrum, has also shown convincing agreement with the NG string~\cite{Athenodorou:2009ms}.

The description of the heavy quark-antiquark potential for fixed separation $R$,
using a fully quantized NG string in 2+D$_\perp$-dimensions, was carried by Arvis with the result~\cite{Arvis:1983fp}
\bea
V(R)=\bigg(\sigma_T^2R^2-\frac{D_\perp}{24\alpha^\prime}\bigg)^{\frac 12} \ , \nonumber
\eea
with the string tension  $\sigma_T=1/2\pi\alpha^\prime$ and $\alpha^\prime=l_s^2$.  The large distance expansion of  the NG potential,
yields
\bea
V(R)\approx \sigma_T R-\frac{\pi D_\perp}{24 R}-\frac{\pi^2}{2\sigma_T R^3}\bigg(\frac{D_\perp}{24}\bigg)^2 \ ,
\eea
the linear confining potential, plus the universal Luscher correction~\cite{Luscher:1980fr} and the Luscher-Weisz correction~\cite{Luscher:2004ib}.
These three contributions are well reproduced by the high precision QCD lattice analysis of the inter-quark
potential  in~\cite{Brandt:2017yzw}. Yet, the full
NG potential becomes imaginary below a critical distance
\bea
\label{RCRIT}
R_c=\pi \sqrt{\frac{D_\perp\alpha^\prime}6}=\frac 12 \frac{|M_0|}{\sigma_T}\rightarrow \frac 13\,{\rm fm} \ .\nonumber
\eea
The  rightmost numerical estimate following  from  $D_\perp=2$, with $\alpha^\prime=1/2m_\rho^2$ the Regge rho meson slope.
This  behavior is tied to the tachyon with negative squared mass $M_0^2<0$ in the dual spectrum,
and  doomed the NG string as a fundamental string in 4-dimensions. Yet the tachyon mass
is also at the origin of the measured large distance $1/R$ and $1/R^3$ corrections exhibited by the QCD string, which is fuzzy and
not fundamental. The inter-quark potential below the critical $R_c$ in QCD, is non-confining.  The lattice results~\cite{Brandt:2017yzw,Brandt:2021kvt}
(see Figs. 1-3 therein) set the deviations at  small separations $r< r_0$, with the Sommer scale $r_0\sim 0.5$ fm.
Hence, the use of the NG string to model QCD interactions, should prove useful away from criticality,  for separations $r>{r_0}>R_c$.
The additional corrections to the NG string stemming from the constraints of Lorentz symmetry~\cite{Aharony:2010cx} are observed to be small in this range.
In a way, this is consistent with the empirical success of the Cornell potential for charmonia~\cite{Eichten:1978tg}.
In what follows, we will make use of these observations.

More specifically,  we will  use the NG string to analyse the interaction potential and scattering amplitude in the Regge limit ($s\gg -t\sim \Lambda_{\rm QCD}^2$), between two QCD
dipoles using the exchange of a NG string in flat 4-dimensional space, without recourse to holography.
When formulated in Euclidean signature, the two calculations parallel each
other with the scattering following from the potential with an Euclidean-angle valued rotation, followed by a pertinent analytical continuation
to the hyperbolic angle or rapidity. This construction was initially suggested   for a pair of quark scattering in~\cite{Meggiolaro:1995cu}.
For completeness, we recall that this stringy
approach to scattering in QCD,
was initially proposed in higher dimensions using the gauge-gravity duality~\cite{Rho:1999jm,Janik:2000aj,Polchinski:2001tt},
and since then discussed  by many e.g.~\cite{Polchinski:2002jw,Brower:2006ea,Brower:2003dtv,Amorim:2021gat} (and references therein).

The present work relies on preceding constructions~\cite{Basar:2012jb,Stoffers:2012zw}, to derive a number of new results:
1/ The static dipole-dipole potential is dominated by the exchange of NG closed strings,
with the tachyon dominant at large distances; 2/ The NG exchanges are characterized by an emergent entropy,
that undergoes a phase-like transition with varying separation; 3/ The scattering of
two dipoles at large rapidity, is dominated by the two-particle irreducible (2PI) NG surface of genus 2,
with a total cross section that is in good agreement with the recently reported data at the LHC;
4/ The NG estimate  of the rapidity and parton-$x$ at saturation; 5/ The contribution of the NG worldsheet fermions to both the
Pomeron intercept, and quantum entanglement at low parton-$x$; 6/ A new entanglement entropy
for multiple NG exchanges in the process of shadowing, that asymptotes a single qubit at the Froissart bound;
7/ The generalization of the NG tachyon diffusion from flat to curved hyperbolic and confining space,
to include evolution in the size of the probing dipoles.

The organization of the paper is as follows: In section~\ref{SEC_I} we detail the construction of the potential
between two static dipoles of fixed and equal size, via the exchange of closed NG strings. The potential is found
to be attractive at all separations, with the NG tachyon dominant at large separations. In section~\ref{SEC_II}
we extend the potential analysis to the scattering amplitude, through a simple rotation in Euclidean signature,
followed by an analytical continuation to rapidity in Minkowski signature. The NG tachyon is shown to dominate the
scattering amplitude at large rapidities. Both the potential and scattering amplitude exponentiate through 2PI ``webs'' that are identified with the  exchange
of a NG string, in leading order in $1/N_c$. This resummation in the scattering channel, yields to saturation of the
total cross section by quantum shadowing, even though the string bits density keeps increasing exponentially
at large rapidities.
 In section~\ref{SEC_V} we review and extend
the string results for quantum entanglement for hadron-hadron in the Regge limit, and DIS scattering in the
low-x regime. We suggest that the quantum entanglement entropy saturates at the Froissart bound.
In section~\ref{SEC_IV} we show how to extend the NG tachyon diffusion in
curved transverse space, by including the size of the probe dipoles  and enforcing conformal symmetry.
The result is reminiscent of the BFKL evolution result following from Mueller$^\prime$s wavefunction
evolution for the total cross section in perturbative QCD. Our conclusions are in
section~\ref{SEC_VI}. A short parallel between the NG tachyon results and pQCD in the Regge limit, is
discussed in the Appendix.

\section{Stringy  dipole-dipole interaction}
\label{SEC_I}
There are many indications from lattice simulations that flux tubes in quenched QCD, can be described by
an effective theory of strings of which the Nambu-Goto (NG) action is the leading contribution~\cite{Kuti:2005xg}.
Remarkably, the NG string appears to describe remarkably well the fuzzy QCD string, even for relatively short
distances. We now use this lattice observation, to analyze the potential between a pair of static dipoles,
in quenched QCD. This potential is amenable to a measurement on the lattice.

This construction parallels closely the analysis of the scattering of a pair of light-like dipoles discussed in~\cite{Basar:2012jb}
(and references therein),
using the holographic construction. To our knowledge, the potential between two parallel dipoles was never analyzed
using the holographic string. It cannot be deduced from the twisted dipoles in~\cite{Basar:2012jb} by taking the twisting angle to
zero, since the  twisted exchange presents an essential singularity in this limit. The singularity reflects on the existence
of a pair creation process through a string-like Schwinger mechanism in the twisted configuration, that is absent in the
untwisted configuration for the potential analysis.

\subsection{Dipole-dipole correlator}
Consider the
static correlators of two identical Wilson loops
\bea
\label{WWPOT}
{\bf WW}=\frac{\langle {\bf W}(a, {\bf b}_\perp){\bf W}(a, {\bf 0}_\perp)\rangle}{\langle{\bf W}\rangle\langle{\bf W}\rangle} \ ,
\eea
with
\bea
\label{WLOOP}
{\bf W}(a, {\bf x}_\perp)=\frac 1{N_c}Tr\bigg({\bf P}{\rm exp}\bigg(ig_Y\int_{{\cal C}_a}d\tau {\bf A}.{\bf v}\bigg)\bigg) \ .
\eea
The contour ${\cal C}_a$ runs along the rectangular loop of side $a$, located at ${\bf x}_\perp$,
with infinite extent along the temporal direction in the v-direction. As usual, the Wilson-loop correlator exponentiates through 2PI ``webs''~\cite{Gatheral:1983cz,Frenkel:1984pz} ,  which we identify to leading order as a closed NG string with genus 2. More specifically,
\bea
\label{WWP}
\ln {\bf WW}\equiv{ \bf WW}_{\rm 2PI}=g_s^2\int \frac{dT}{2T}\,{\bf K}(T) \ ,
\eea
with $g_s\sim \frac 1{N_c}$ the string coupling, and
\bea
\label{PROP}
{\bf K}(T)=\int_TD[x]\,e^{-S[x]+{\rm ghost}} \ ,
\eea
the NG string partition function with cylindrical topology and modulus $T$.
The NG action in conformal gauge is~\cite{Polyakov:1986cs}
\bea\la{action}
S[x] &=& \frac{\sigma_T}{2} \int_0^T d \tau \int_0^1 d \sigma ~ (\dot{x}^\mu \dot{x}_\mu + {x'}^\mu {x'}_\mu ) \ .
\eea
Here $\dot{x}=\partial_\tau x$ and $x^\prime=\partial_\sigma x$. The string tension is $\sigma_T=1/(2\pi \alpha^\prime)$.

The evaluation of  (\ref{WWP}) in string theory, is in general difficult due to the finite dipole sizes, and we need to make reasonable
approximations. For small size dipoles, we will assume that the cylindrical boundaries are highly pinched, and approximate them by
straight lines. The exchanged closed string forms a funnel, with linear end-points, much like the exchange between static D0 branes,
as detailed for the twisted dipoles in~\cite{Basar:2012jb}. However, funnels with higher windings or N-ality are not suppressed between D0 branes,
but are suppressed between dipoles of finite transverse size. A physical interpretation of the final result, will allow for a simple extraction
of the dipole-dipole  potential from this approximation  below.

\subsection{Static dipole-dipole potential}

With this in mind,  we now decompose the string embedding coordinates using the worldsheet normal modes in $2+D_\perp$ flat space,
with linear  and periodic boundary conditions in the affine time with period $T$~\cite{Qian:2014jna}
\bea\la{untwistdecompose}
x^0 (\tau, \sigma) &=&  X + \frac{cW}{\sigma_T} \tau +\sum_{m=-\infty}^\infty \sum_{n=1}^{\infty} x_{m,n}^0 \exp\bigg({i 2 \pi m \frac{\tau }{T}}\bigg)
\cos (\pi n \sigma) \ ,
\nonumber\\
x^1 (\tau, \sigma) &=& \sum_{m=-\infty}^\infty \sum_{n=1}^{\infty} x_{m,n}^1 \exp\bigg({i 2 \pi m \frac{\tau }{T}}\bigg) \sin (\pi n \sigma) \ , \nonumber\\
x_\perp (\tau, \sigma) &=& \bigg(\sigma - \frac{1}{2}\bigg) b_\perp + \sum_{m=-\infty}^\infty \sum_{n=1}^{\infty} x_{m,n}^\perp \exp\bigg({i 2 \pi m \frac{\tau }{T}}\bigg) \sin (\pi n \sigma) \ .
\eea
with $c=1/{l_s}$. Since
\be
x^0 (\tau+T, \sigma) -x^0 (\tau , \sigma) =W \bigg(\frac{T}{\sigma_T l_s}\bigg) \ ,
\ee
we interpret $W=0, \pm 1, \pm 2, ...$ as a winding number, with $2\pi l_s T$  the circumference of the cylindrical funnel.
Below we will show that this is at the origin of an effective temperature for the exchanged closed strings,
in the dipole-dipole potential. $X,W$ plays the role of collective coordinates.

 In terms of (\ref{untwistdecompose}),
all integrals in (\ref{PROP}) are Gaussian with the result
\be\la{untwistfreepropagator}
\int dX\sum_{W} \bold{K}(T , W) = \frac{a^2 X}{l_s^3}
\sum_{W}\exp\left( - \frac{T}2 \bigg[\sigma_T b^2
+ \frac{ c^2W^2}{\sigma_T T^2} \bigg]\right) \left[\prod_{n=1}^\infty 2 \sinh \left( \frac{n \pi T}{2} \right) \right]^{-D_\perp} \ .
\ee
The diverging products can be regularized by standard zeta function regularization, using the representation
\be
\sinh (\pi x) = \pi x \prod_{m=1}^\infty \left( 1 + \frac{x^2}{m^2} \right) \ .
\ee
With this in mind, and trading the re-summation over the windings using the Poisson summation formula, we obtain
\bea\la{windeq1}
\int dX \sum_W \bold{K} (T, W)
=\sqrt{\frac{1}{c^2\alpha^\prime T }} \frac{a^2X}{l_s^3} \sum_{k}
\exp\left(- \frac{T}2\bigg[\sigma_T b^2 +\frac{2\pi k^2}{c^2\alpha^\prime T^2}\bigg] \right)
\eta^{- D_\perp} \left( i \frac{T}{2} \right) \ , \nonumber \\
\eea
where $\eta(x)$ is Dedekind eta function
\bea
\label{ETA}
\eta(\tau)=q^{\frac 1{24}}\prod_{n=1}^{\infty}(1-q^n)\qquad q=e^{2i\pi\tau}
\eea
Note that (\ref{ETA}) satisfies $\eta (i x) = \eta (i/x) /\sqrt{x}$, and relates to the string density of modes~\cite{Fubini:1969wp}
\bea
\label{DEDE}
\eta^{- D_\perp} \big(ix\big) =
 x^{\frac{D_\perp}{2}} e^{\frac{ \pi D_\perp }{12 x }} \sum_{n=0}^\infty d(n) e^{-n \frac{2 \pi}{x}} \ ,
\eea
with $d(n)$ being the string density of states normalized to $d(0)=1$, with asymptotically
\bea
\label{DENSITY}
d(n)\approx  C\,n^{-\frac {D_\perp+3}4}\,{e^{2\pi\sqrt{D_\perp n/6}}} \ .
\eea
The Poisson re-summation trades the sum over the windings $W$ of the closed string exchanges,
with the dual sum over the N-alities or $k$ fluxes~\cite{Basar:2012jb}. For source dipoles as Wilson loops in the fundamental representation,
$k=1,..,[\frac{N_c}2]$ which runs to infinity in the large $N_c$ limit. For QCD with $[\frac 32]=1$,
only the $k=1$ N-ality is to be retained.

The leading contribution to the static  potential for two parallel dipoles of size $a$, is given by the
2-particle irreducible (2PI) string exchange
\bea
\label{FFXX}
V_{DD}(b)=-\frac{{\rm ln}\bold{WW}}{X}=-g_0^2\sum_{n=0}^\infty d(n)\Delta(m_n, b)
\eea
with the exchanged scalar propagator
\bea
\label{PROX}
\Delta(m_n, b)=\int \frac{dk^{D_\perp+1}}{(2\pi)^{D_\perp+1}}\frac{e^{ik\cdot b}}{k^2+m_n^2}=
\frac 1{2\pi}\bigg(\frac{m_n}{2\pi b}\bigg)^{\frac {D_\perp-1}2}\,K_{\frac {D_\perp-1}2}(m_nb)
\eea
and coupling to the dipole $g_0=\frac{g_s a}{\alpha^\prime}(\sqrt{2\pi}\alpha^\prime)^{\frac{D_\perp}4}$.
Here $\bold{K}_\alpha (x)$ is the modified Bessel function, and
$d(n)$ is the canonical string density of states with $d(0)=1$, and $g_s$ the string coupling.
(\ref{FFXX}) amounts to a tower of closed string exchanges or glueballs, with radial masses
\bea
\label{MASSN}
 m_{n}=\frac{1}{c\alpha^\prime}\ \left(1- \frac{ D_\perp c^2}{12  \pi \sigma_T}+ \frac{2 n \pi c^2}{\pi^2 \sigma_T} \right)^{\frac 12}=
\sigma_T\beta\left(1-\frac{\beta_H^2}{\beta^2}+\frac{8\pi n}{\sigma_T\beta^2}\right)^{\frac 12} \ ,
\eea
and with the inverse Hagedorn temperature $\beta_H=\sqrt{\pi D_\perp/3\sigma_T}$.
$\beta=2\pi/c=2\pi l_s$  is the circumference of the exchanged cylindrical worldsheet.

For large $b$, the exchange in (\ref{FFXX}) is dominated by the tachyon mode with $n=0$
\bea
\label{TACHO}
m_{0}=\sigma_T\beta\left(1-\frac{\beta_H^2}{\beta^2}\right)^{\frac 12}=\frac 1{l_s}\left(1-\frac{D_\perp}6\right)^{\frac 12} \ ,
\eea
which  is still real positive for $D_\perp<6$. For $D_\perp=2$ and using half the rho meson Regge slope
$\alpha^\prime =1/4m_\rho^2$ for a closed string, we have $m_0\approx 1257$ MeV, which is close to the $m_{0^{++}}=1475$ MeV glueball
 reported on the lattice~\cite{Meyer:2004gx}.
The attractive and static dipole-dipole potential
follows as
\bea
\label{VLARGE}
V_{DD}(b)\approx -\frac{g_s^2 a^2\sqrt{\pi \sigma_T}}{ \alpha^\prime}
\left(\frac{\pi\alpha^\prime m_{0}}{b}\right)^{\frac{D_\perp-1}2}
\bold{K}_{\frac{D_\perp-1}{2}} \left( m_{0}b\right) \ .
\eea
For short separations, the exchange is also attractive
\bea
\label{VSHORT}
V_{DD}(b)\approx -\bigg(\frac{g_s^2 a^2\sqrt{\pi \sigma_T}}{ 2\alpha^\prime}\sum_n d(n)\bigg)
\Gamma\bigg(\frac{D_\perp-1}2\bigg)\,\bigg(\frac 1{\sigma_T b^2}\bigg)^{\frac{D_\perp-1}2} \ ,
\eea
and Coulombic for $D_\perp=2$, i.e. $V_{DD}(b)\sim -g_s^2/b$. However, this contribution
signals the onset of a critical NG string, with a diverging mode sum for the overall coefficient.
At short separations, the exchange is not confining. It is dominated by 2-gluon Coulomb exchange in the $0^{++}$ channel.
The Casimir-Polder contribution characterizes the fully non-confining potential at large separations~\cite{Shuryak:2000df}
(and references therein).

Recall that the string coupling is $g_s=f(\lambda)/N_c$,
with $f(\lambda)$ a non-universal function of the large $^\prime$t Hooft coupling $\lambda=g_Y^2N_c$.
For instance, in  holographic models,
$f(\lambda)=\lambda/4\pi$ (${\cal N}$=4 SUSY) and $f(\lambda)=(\lambda/3)^{\frac 32}/\pi$
(Witten model).  This observation shows that the 2-PI contribution (\ref{FFXX}) is dominant
in large $N_c$, as the higher \#-PI contributions are suppressed by $(1/N_c^2)^{1+\#}$.

\subsection{Emergent entropy in dipole-dipole interaction}

The circumference $\beta=2\pi l_s$ in (\ref{FFXX}),
plays the role of an inverse effective temperature, associated with the
spatial exchange of closed strings or glueballs,  in the transverse b-direction.
Remarkably, $T_R=1/\beta=1/2\pi l_s$ is identical to the Rindler temperature of
falling matter on the stretched horizon, a membrane a string length away from the
event horizon of a stationary black-hole. With this in mind, we may interpret
the potential in (\ref{FFXX}) as a {\it free energy}. As a result, the
 stringy exchange  in the dipole-dipole potential, carries an emergent entropy,
\bea
\label{SEV}
S_{EV}(b)= \beta^2\bigg(\frac{\partial V_{DD}(b)}{\partial \beta}\bigg)
\eea

To understand the nature of this entropy and how it may relate to the spatial entanglement entropy in QCD4 at large $N_c$, we note that (\ref{FFXX}) at large separation is given by
\bea
\label{EEX1}
V_{DD}(b)\sim \sqrt{\frac \pi 2} \bigg(\frac{g_s^2a^2}{{\alpha^\prime}^{\frac 32}}\bigg)\bigg(\frac {2\pi\alpha^\prime}{b^2}\bigg)^{\frac{D_\perp-1}2}
\sum_{n=0}^\infty d(n)\,x_n^{\frac {D_\perp}2-1}\,e^{-x_n}
\eea
with $\alpha^\prime=l_s^2$ and $x_n=m_n b$. Inserting (\ref{EEX1}) in (\ref{SEV}) gives
\bea
\label{EEX2}
S_{EV}(b)\sim  (2\pi)^{\frac{D_\perp}2+4}\bigg(\frac{g_s^2a^2}{2b^2}\bigg)\bigg(\frac{l_s}b\bigg)^{D_\perp-1}\,
\sum_{n=0}^\infty d(n)\,x_n^{\frac {D_\perp}2-2}\,e^{-x_n}
\eea
(\ref{EEX2})  amounts to the entropic function at large $x_n\gg 1$
\bea
\label{EEX3}
C_{DD}(b)=\frac{\partial S_{EV}(b)}{\partial{\rm ln}b}\sim  - (2\pi)^{\frac{D_\perp}2+4}\bigg(\frac{g_s^2a^2}{2b^2}\bigg)\bigg(\frac{l_s}b\bigg)^{D_\perp-1}\,
\sum_{n=0}^\infty d(n)\,x_n^{\frac {D_\perp}2-1}\,e^{-x_n}
\eea
Using the radial mass spectrum (\ref{MASSN}) and the string density of states (\ref{DENSITY}), the sum is dominated by the large-n contribution
\bea
\label{EEX4}
\sum_{n=0}^\infty d(n)\,x_n^{\frac {D_\perp}2-1}\,e^{-x_n}\sim \int_{n_{\rm low}}^\infty dn\, n^{-\frac 54}\,e^{-m_n b+2\pi\sqrt{\frac{D_\perp n}6}}
\eea
Using the change of variable $t=\sqrt{8\pi n/\sigma_T}/\beta$ we can rewrite  (\ref{EEX4}) as
\bea
\int_{t_{\rm low}}^\infty \frac {dt}{t^{\frac 32}}\,e^{\frac 12 \sigma_T\beta \beta_H t-\sigma_T\beta b\sqrt{1+t^2}}
\eea
which can be undone by saddle point approximation for $\beta/b\ll 1$, with the result
\bea
\label{EEX4X}
\sum_{n=0}^\infty d(n)\,x_n^{\frac {D_\perp}2-1}\,e^{-x_n}\sim e^{-\sigma_T b\sqrt{b^2-(\beta_H/2)^2}}
\eea
to exponential accuracy.
The branch point singularity $b=\frac 12 \beta_H=R_c$ reflects on the diverging sum for small distances in agreement with (\ref{RCRIT}).

We now recall that the entropic function for spatial entanglement in QCD4 at large $N_c$, was argued to be of  the form~\cite{Klebanov:2007ws}
\bea
\label{EEX5}
C_{1+3}(b)\sim \frac 1{32\sqrt\pi}\,\frac {V_2}{b^2}\,\sum_{n=0}^\infty d(n)(M_n b)^{\frac 32}e^{-2M_nb}
\sim  \int^\infty dM \,M^\alpha e^{(\beta_H-2b)M}
\eea
where $M_n\sim \sqrt{n}/\alpha^\prime$ was used   for the asymptotic of the glueball spectrum. If instead, the exact radial glueball spectrum
(\ref{MASSN}) with $M_n=\frac 12 m_n$ which carries the same asymptotics is used in (\ref{EEX4}), then to exponential accuracy in the entropic
function for spatial entanglement in QCD4
\bea
\label{EEX6}
\sum_{n=0}^\infty d(n)(M_n b)^{\frac 32}e^{-2M_nb}\sim  e^{-\sigma_T b\sqrt{b^2-(\beta_H/2)^2}}
\eea
in agreement with (\ref{EEX4X}).

We conclude that the entropic function constructed from the emergent entropy (\ref{SEV}) associated to two interacting dipoles,
shares the same asymptotic behavior as the entropic function
for spatial entanglement in QCD4 at large $N_c$. It also diverges at the same location $b=\frac 12 \beta_H$.  The divergence reflects on
the transition from a confining to a deconfining phase  as probed by  (\ref{SEV}) and similarly by (\ref{EEX6}). Amusingly, the mode sums in both cases are identical for
$D_\perp=5$ or  a NG in 2+5-dimension, somehow reminiscent of the holographic proposal.


\section{Stringy dipole-dipole scattering}
\label{SEC_II}
At large center of mass energy with a large rapidity gap, the hadron-hadron scattering amplitude is universal.
The  amplitude is that of two fixed size dipoles scattering elastically. For small angle scattering, the amplitude
is dominated by gluon exchanges with vacuum quantum numbers. In perturbative QCD, the exchange is
captured by the BFKL resummation of rapidity ordered gluons, the so-called hard Pomeron. In non-perturbative
QCD, the resummation is captured by Reggeized gluons. In the planar approximation, the exchange is string-like
with the topology of a cylinder, the so-called soft Pomeron. The existence of a  hard and soft Pomeron at
large rapidity,  was initially pointed out in~\cite{Donnachie:1998gm}. It finds a natural description  in the gravity
dual approach to QCD~\cite{Polchinski:2002jw,Brower:2006ea},  using a critical  string in 10 dimensions  in the
conformal limit (hard Pomeron),  followed by conformal symmetry breaking  (soft Pomeron).

The purpose of this section is to show that the NG string which is non-critical, allows for the description of the
elastic dipole-dipole scattering amplitude using $1/N_c$ counting rules, already in 4 dimensions. The result is
a soft Pomeron with parameters that are distinct from the holographic results. The scattering construction
parallels that of the dipole-dipole interaction, showing the inter-connecteness of the potential and scattering
problems. The hard Pomeron can also be retrieved, by allowing the tachyon mode in the NG string to diffuse
both in transverse and longitudinal size, a point inspired by holography~\cite{Stoffers:2012zw}. Since the NG
string provides the closest description of the QCD string potential, it should prove relevant for the QCD scattering
amplitude in the eikonal limit. In particular, a more transparent approach to the unitarization of the cross section,
as well as the partonic string bit content and entanglement, are seen to emerge.

The present construction using a NG string in $D=4$ dimensions is the closest to the holographic construction
using $D>4$ presented in~\cite{Stoffers:2012zw,Basar:2012jb}, with an essential difference in the Pomeron
intercept. The twisted dipole-dipole correlator follows from the reduction of the eikonal QCD scattering amplitude,
and then evaluated using the closed and non-critical  NG string in flat $D=4$ dimensions. The exchange is solely
in the confining regime of QCD.  It differs from the holographic Pomeron discussed
 in~\cite{Polchinski:2002jw}, by the value of its intercept, the coupling to the sourcing dipoles, and the absence of a
 conformal limit. This can remedied as we discuss below.

\subsection{Scattering amplitude}
In Euclidean signature, the scattering between two dipoles follows the same analysis as that for the potential
between two static dipoles presented above, with one difference: the dipoles are not palarell but slated at
an angle $\theta$. This angle maps by analytical continuation to the rapidity or boost angle $\chi$, thanks to Minkowski
historical observation. With this in mind, a  rerun of the preceding arguments gives  for the twisted worldsheet propagator~\cite{Basar:2012jb}
\be
\la{freepropagator}
\bold{K} (T, \theta) = \frac{a^2}{\alpha'} \frac{e^{- \frac{\sigma_T}{2} T b^2} }{ 2 \sinh \left(\frac{\theta T}{2}\right) }
\prod_{n=1}^\infty \prod_{s =\pm} \frac{\sinh \left( \frac{n \pi T}{2} \right) }{\sinh \left[ \frac{T (n \pi + s \theta)}{2} \right] }
\left[\prod_{n=1}^\infty 2 \sinh \left( \frac{n \pi T}{2} \right) \right]^{-D_\perp} \ .
\ee
The details regarding the twisted string worldsheet mode decomposition analogous to (\ref{untwistdecompose}), followed by the
detailed mode integration
leading to (\ref{freepropagator}) are given in~\cite{Basar:2012jb}.
The double analytical continuation $T\rightarrow iT$ and
 $\theta\rightarrow  - i \chi$,  maps this twisted  dipole-dipole correlator onto the
 scattering amplitude of two light-like Wilson loops. With this in mind, inserting (\ref{freepropagator}) in the
 corresponding ${\bf WW}$ 2PI correlator and using (\ref{ETA},\ref{DEDE}),  yield~\cite{Basar:2012jb}
\bea\la{ww1}
\bold{WW}_{\rm 2PI}(\chi,a, b) &=& \frac{ g_s^2 a^2}{4 \alpha'} \sum_{k = 1}^\infty \frac{(-1)^k}{ k } e^{- k\frac{\pi \sigma_T \bold{b}^2}{\chi} } \eta^{- D_\perp} \bigg(\frac{i k \pi }{\chi}\bigg) \nonumber\\
&=& \frac{ g_s^2 a^2}{4 \alpha'} \sum_{k = 1}^\infty\sum_{n=0}^{\infty} \,d(n)\,\frac{(-1)^k}{ k } \bigg(\frac{k\pi}{\chi}\bigg)^{\frac {D_\perp}2}
{\rm exp}\bigg(-\frac{2\chi}k \bigg[n+\frac{\bold{b}^2}{\alpha^\prime(2\chi/k)^2} -\frac{D_\perp}{24}\bigg]\bigg) \ , \nonumber\\
\eea
after analytical continuation, with $\chi \approx \ln (\alpha^\prime s)$ identified as the rapidity, for large invariant mass $\sqrt s$.
The scattering amplitude in momentum space is
\bea\la{scatteringamplitude}
\frac{1}{- 2 i s} \mathcal{T}_{DD} (\chi, q) &\approx& \int d^2 \bold{b}~ e^{i \bold{q}_\perp \cdot \bold{b} }\,  \bold{WW}_{\rm 2PI}(\chi, a,b)\nonumber\\
&\approx& \frac{\pi^2 g_s^2a^2}{2} \sum_{n=0}^\infty \sum_{k=1}^\infty d(n)\frac{(-1)^k }{k} \left(\frac{k \pi}{\chi}\right)^{\frac{D_\perp - 2}{2}}
\exp \left(-\frac{2\chi}k\left[n+\frac{\alpha^\prime}4 {\bf q}_\perp^2-\frac{D_\perp}{24}\right]\right) \ . \nonumber\\
\eea
Again, $k$ sums over the N-ality with $k=1,..,[\frac{N_c}2]$ all the way to infinity at large $N_c$. In our case, only the
$k=1$ term contributes to the scattering of two dipoles as twisted Wilson loops, in the fundamental representation of SU$(3_c)$.
With this in mind, and in  the large rapidity limit, (\ref{scatteringamplitude}) simplifies
\bea\la{scatteringamplitudeX}
\mathcal{T}_{DD} (\chi, q) \approx is \,(\pi g_s a)^2\,\left(\frac \pi\chi\right)^{\frac {D_\perp}2-1}
\exp \left(-\chi\left[\frac{\alpha^\prime}2\bigg( {\bf q}_\perp^2+M_0^2\bigg)\right]\right) \ ,
\eea
with the tachyon squared mass $ M_0^2=-\frac{D_\perp}{6\alpha^\prime}$.

The closed string exchange amounts to a Pomeron exchange, with a Regge trajectory
\bea
\label{INTER}
\alpha_{\mathbb P}(t)=\frac{D_\perp}{12}+\frac{\alpha^\prime}2 t \ ,
\eea
hence a dipole-dipole (hadron-hadron) scattering amplitude that rises as $\sigma_{\mathbb P}(s)\sim s^{\alpha_{\mathbb P}(t)}$.
In the Regge limit with $-t\ll s$,  this amplitude  is dominated by a single NG string exchange,  given by
\bea
\label{HAD1}
{\cal A}(s, t )\sim -2is\int d^{2}be^{iq\cdot b}\,{\bf WW}_{\rm 2PI}(s, a, b)\sim is^{1+\alpha_{\mathbb P}(t)} \ ,
\eea
with $t=-q^2$.

\subsection{Cross section and Froissart bound}

The elastic scattering amplitude (\ref{HAD2}) yields the total cross section
\bea
\label{HAD2}
\sigma(s)=\frac 1s {\rm Im}\, {\cal A}(s, 0)\sim -2\int d^{2}b\,{\bf WW}_{\rm 2PI}(s, a,b)\sim s^{\alpha_{\mathbb P}(0)} \ ,
\eea
by the optical theorem. (\ref{HAD2})  increases  with the squared invariant mass $s$, in violation of unitarity.
This shortcoming can be addressed by noting that $\langle {\bf WW}\rangle$ as a correlator of two Wilson loops, requires
the exponentiation of all the 2PI contributions in leading order in $\frac 1{N_c}$, much like in the potential between the two static dipoles discussed earlier
\begin{align}
{\bf WW}(\chi,a,b)=\frac{\langle {\bf W}_{\frac{\chi}{2}}(a, {\bf b}_\perp){\bf W}_{-\frac{\chi}{2}}(a, {\bf 0}_\perp)\rangle}{\langle{\bf W}\rangle\langle{\bf W}\rangle}=\exp \bigg[{\bf WW}_{\rm 2PI}(\chi,a,b)\bigg] \ ,
\end{align}
${\bf WW}_{\rm 2PI}$ is the 2PI ``web'' contributions~\cite{Gatheral:1983cz,Frenkel:1984pz}, which is dominated by a string exchange with genus 2  as detailed above, with
higher geni contributions  suppressed by powers of  $g_s^2\sim 1/N^2_c$. Since ${\bf WW}\equiv {\cal S}$ identifies with the full $S$-matrix, and using  ${\cal S}=1+i{\cal T}$ as detailed in Appendix~\ref{APP}, we obtain
\bea
\label{HAD3}
\sigma(s)=2\int d^2b \ {\rm Re}\big(1-{\cal S}(\chi,a,b)\big)= 2\int d^{2}b\,\big(1-e^{{\bf WW}_{\rm 2PI}(\chi, a,b)}\big) \ ,
\eea
since ${\bf WW}_{\rm 2PI}(\chi, a,b)$ is real. To proceed, it  is useful to recast the tachyon contribution in (\ref{ww1}) in impact parameter space, as follows~\cite{Basar:2012jb}
\bea\la{ww1X}
\bold{WW}_{\rm 2PI}(\chi,a, b) \approx
- \frac{ g_s^2 a^2}{4 \alpha'} \bigg(\frac{\pi}{\chi}\bigg)^{\frac {D_\perp}2}\,e^{-S_{\rm cl}-S_{\rm 1loop}} \ ,
\eea
The first contribution in the exponent of (\ref{ww1X})
\bea
\label{WW2}
S_{\rm cl}=\sigma_T \int_0^{T_P}{\rm cos}^2(\chi\tau)\, bd\tau\int_0^1 bd\sigma=\frac 12 \sigma_T\beta b=\frac  {b^2}{2\alpha^\prime \chi} \ ,
\eea
is identified with a semi-classical worldsheet instanton,  with a tunneling  time $T_P=1/\beta=2\pi b/\chi$. The second contribution in the exponent of (\ref{ww1X})
\bea
\label{WW3}
S_{\rm 1loop}=\frac{D_\perp}2{\rm lndet}(-\partial_\perp^2)=-\frac{\pi D_\perp}6 \frac {b}{\beta}=-\frac {D_\perp}{12}\chi \ ,
\eea
is the  1-loop zeta regulated corrective action, around the worldsheet instanton.

\begin{figure}[!htb]
\includegraphics[height=6cm]{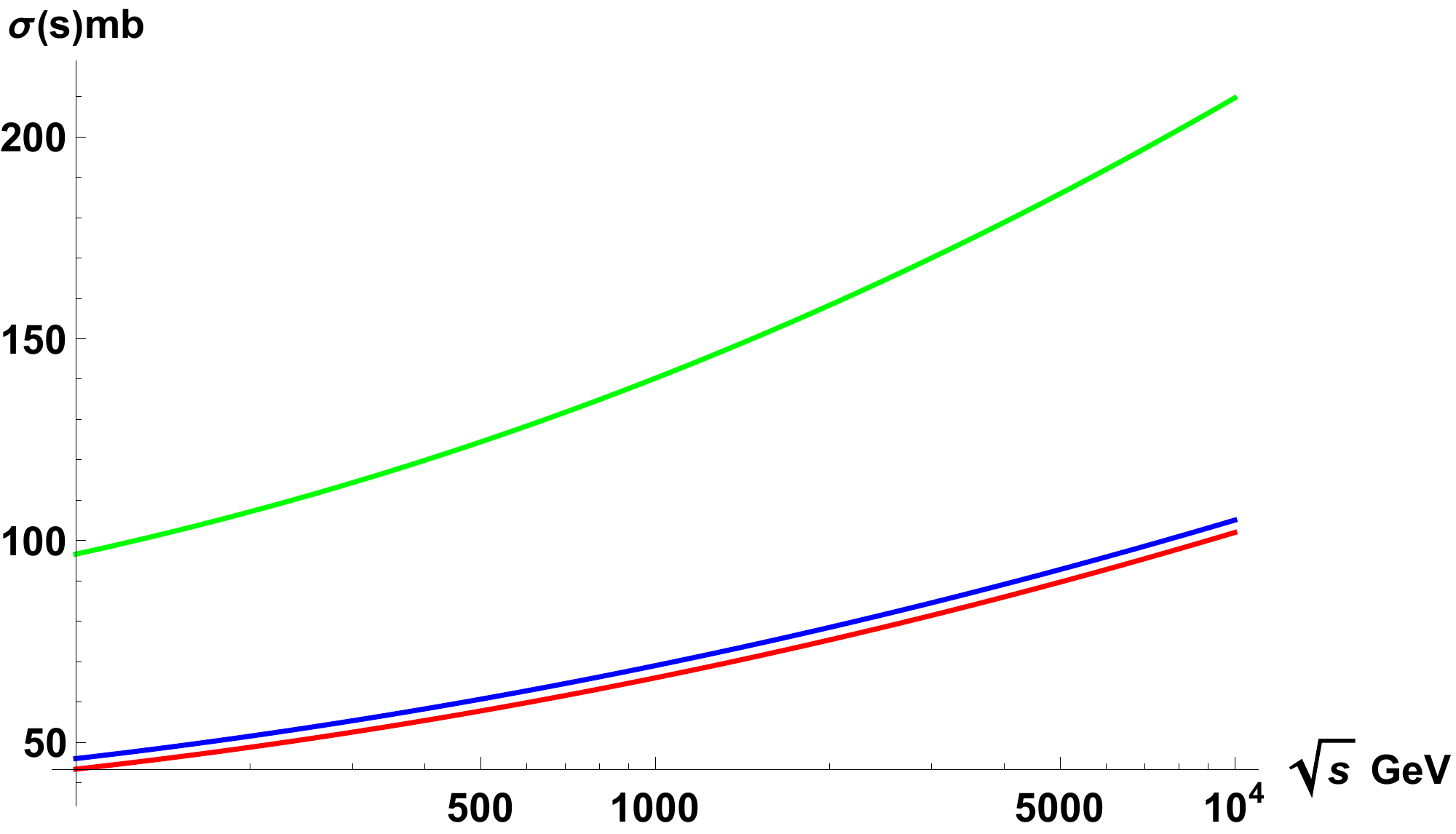}
 \caption{Total cross section in mb versus $\sqrt{s}$ in GeV: solid-red-lower curve is the empirical $pp$ cross section as parametrized
 by the COMPETE collaboration in~\cite{COMPETE:2002jcr}  and quoted in  (\ref{DATA}). It is in agreement with the cross sections
 currently measured by the LHC~\cite{TOTEM:2017asr}.
 The solid-blue-lower curve is the NG result (\ref{SSX}) with $a=l_s$ and $g_s=1$ and ${\cal O}$=30, with $\alpha^\prime=1/4m_\rho^2$,
 while the  solid-green-upper curve is for $\alpha^\prime=1/2m_\rho^2$. }
  \label{fig:CROSS}
\end{figure}

Using (\ref{ww1X}-\ref{WW3}), we now note that the integrand in (\ref{HAD3}) is controlled by the exponent in (\ref{ww1}),
which is a tradeoff between the minimal worldsheet instanton action $S_{\rm cl}$ and its 1-loop quantum correction $S_{\rm 1loop}$.
The black disc radius is  reached when
\bea
\label{HAD4}
S_{\rm cl}=|S_{\rm 1loop}|\rightarrow b_{\rm max}^2=\frac{D_\perp\alpha^\prime}6 \chi^2 \ ,
\eea
so that $e^{{\bf WW}_{\rm 2PI}}\rightarrow \theta(b-b_{\rm max})$.
As a result (\ref{HAD3}) yields the total cross section
\bea
\label{HAD6}
\sigma (s)\sim 2\int d^{2}b\,\theta (b_{\rm max}-b)\sim 2\pi\,b_{\rm max}^{2}=2\pi\alpha^\prime\frac{D_\perp\,\chi^2}{6} \ .
\eea
At large rapidity, the 2PI NG contribution  saturates the Froissart bound,   with a scale fixed by the string tension
$\sigma_T=1/2\pi\alpha^\prime$,  and not the pion mass as suggested in~\cite{Nussinov:2008nz,Basar:2012jb,Kharzeev:2017azf}.

More specifically, (\ref{HAD3}) evaluates exactly to
\begin{align}
\label{SSX}
\sigma(s)=2\pi \alpha^\prime \bigg(\frac{D_\perp\chi^2}{6}- \chi\ln (D_\perp\chi)+\left(\frac{a^2g_s^2\pi}{4\alpha^\prime}\chi+\gamma_E\right)
+{\cal O}\bigg(e^{-\frac{a^2g_s^2\pi}{4\alpha^\prime\chi}e^{\frac{D_\perp\chi}{6}}}\bigg)\bigg)\ ,
\end{align}
with $\chi={\rm ln}(\alpha^\prime s)$, in agreement with the estimate (\ref{HAD6}).
The new result (\ref{SSX}) stemming from the NG exchange, is to be compared with the empirical
parametrization  of the  $pp$  data by the  COMPETE  collaboration~\cite{COMPETE:2002jcr}
\bea
\label{DATA}
\sigma^{pp}(s)\sim \bigg(35.5 +0.307\,{\rm ln}^2\bigg(\frac s{29.1\,{\rm GeV}^2}\bigg)\bigg)\,{\rm mb} \ ,
\eea
after dropping the Reggeon contributions at large $\sqrt{s}$. In Fig.~\ref{fig:CROSS} we show the NG result for the total cross section
(\ref{SSX}) for $a=l_s$ and $g_s=1$  with ${\cal O}$=30: green-solid-upper curve
with $\alpha^\prime=l_s^2=1/2m_\rho^2$ (the rho meson trajectory slope),  and red-solid-curve with  $\alpha^\prime=l_s^2=1/4m_\rho^2$
(half the rho meson trajectory slope). The empirical
parametrization  (\ref{DATA}) blue-solid-lower curve, has been used by the COMPETE collaboration to reproduce the compiled $pp$ and $p\bar p$ data,
two decades ago. It is in good agreement with the recently reported TOTEM measurements for $pp$
at the highest $\sqrt{s}=13$ TeV at the LHC~\cite{TOTEM:2017asr}. The NG result is mostly sensitive to the string length,
and is undistinguishable from the COMPETE parametrization for $\alpha^\prime=l_s^2=1/4m_\rho^2$. Note that the value of this
parameter is tied to  glueball spectrum in (\ref{TACHO}). This shows the duality between the scattering amplitude of two dipoles,
and the  static potential between two dipoles.

\begin{figure}[!htb]
\includegraphics[height=6cm]{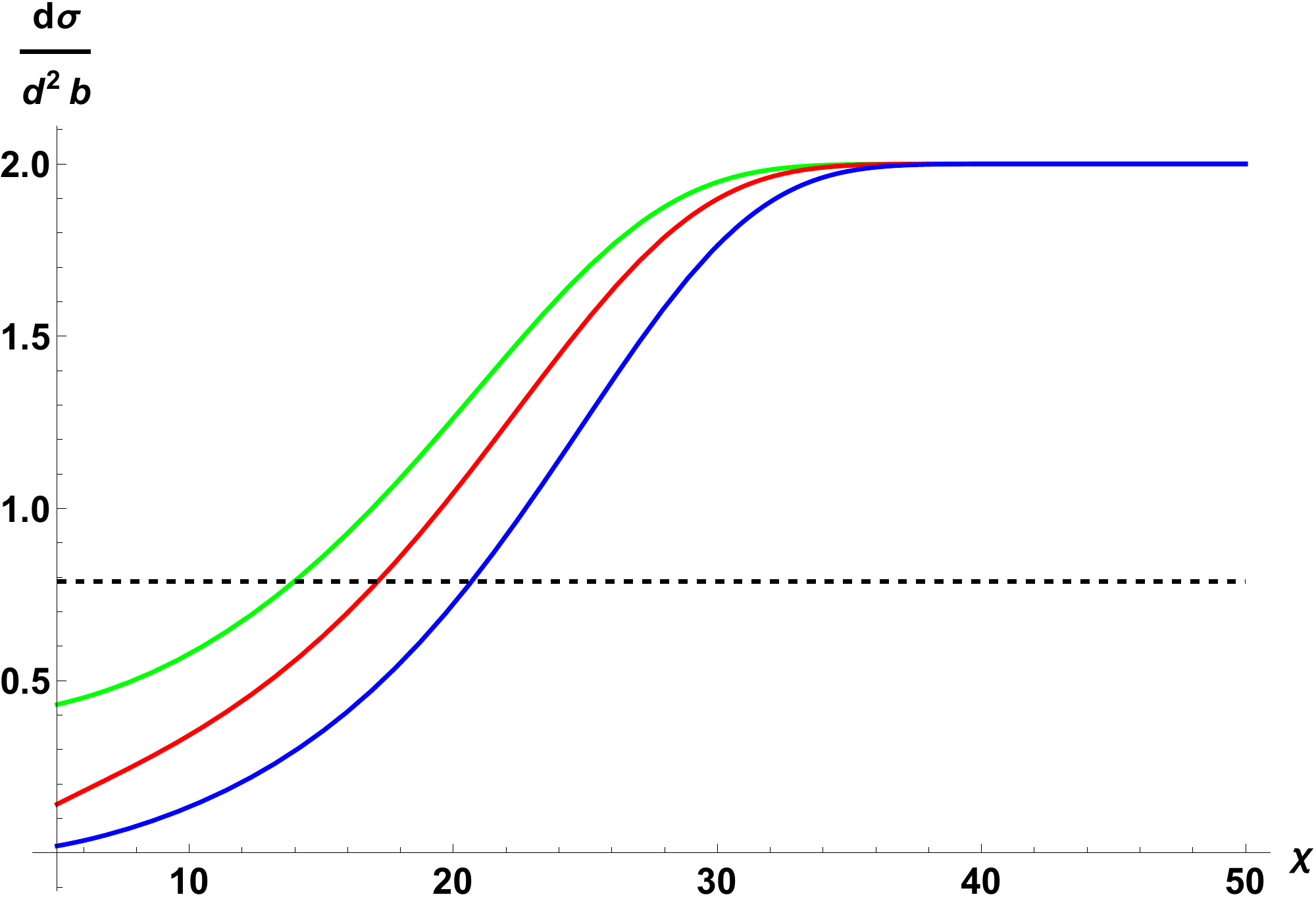}
 \caption{Differential cross section (\ref{HAD3X}) for fixed impact parameter $b$ between two dipoles of fixed size $a=1$ and $g_s=1$,
  as it unitarizes by shadowing, at large rapidity $\chi$.   The  green-solid upper curve, red-solid middle curve and
the blue-solid lower curve are for impact parameters $b=2,4,6$, in units of the string length. The dashed line
follows from  the saturation condition (\ref{XSATX}).}
  \label{fig:SIGMAABX}
\end{figure}

\subsection{Shadowing of wee string bits}
We now note that the total cross section (\ref{HAD3}) amounts to
\bea
\label{HAD3X}
\sigma(s)\sim 2\int d^{2}b\,N(s,a,b) \ ,
\eea
with
\bea
\label{SBX}
\frac 12 \frac{d\sigma}{d^2b}=N(s,a,b)=1-S(s,a,b)=1-e^{{\bf WW}_{\rm 2PI}(s,a,b)} \ .
\eea
$\frac 12$  the {\it effective} number of wee-string-bits  flowing through the cylindrical annulus  $2\pi b db$.
The effective number $N$, obeys a non-linear diffusion-like
equation
\begin{align}
\bigg(\partial_{\xi}+M_0^2-\nabla_\perp^2\bigg)\ln (1-N)=0 \ ,
\end{align}
with $\xi=\frac 12 \alpha^\prime \chi$ playing the role of an effective (Gribov) time, or
\begin{align}
\partial_{\xi} N-M_0^2(1-N)\ln (1-N)-\nabla_\perp^2 N-\frac{(\nabla_\perp N)^2}{1-N}=0 \ .
\end{align}
In the small  number limit  $\ln (1-N)\sim -N$,  one recovers the linear diffusion equation, with
non-linear corrections for larger $N$,  that cause saturation asymptotically.
For instance, in the quadratic approximation,  the non-linear evolution is given by
\begin{align}
\bigg(\partial_{\xi}+M_0^2-\nabla_\perp^2\bigg)N-\frac{M_0^2}{2}N^2-(\nabla_\perp N)^2+{\cal O}(N^3)=0 \ ,
\end{align}
which is  reminiscent of the non-linear  Gribov-Levin-Ryskin equation, for the unintegrated gluon distribution~\cite{Gribov:1983ivg}.

In Fig.~\ref{fig:SIGMAABX} we show the behavior of the integrand in the total cross section (\ref{HAD3X}) versus the rapidity,
for three different values of the impact parameter $b$.
The  green-solid upper curve, red-solid middle curve and
the blue-solid lower curve are for impact parameters $b=2,4,6$, in units of the string length. We have set
the string coupling  $g_s=1$, and the static dipole sizes $a=1$ in units of the string length. The dependence on the
impact parameter is mild.

\subsection{DIS view of wee string bits}


In the stringy approach to the Pomeron and unitarization, the picture of a hadron at large rapidities or small-x,
is different from that following from pQCD, where  a hadron at large rapidity $\chi$ preserves its transverse
size, and shrinks its longitudinal size by the gamma factor $\gamma=e^{\frac 12 \chi}$. In contrast, when a string is exchanged,
the hadron transverse size grows logarithmically as $|\Delta x_\perp|\sim \sqrt{\chi\alpha^\prime}$, while its light front longitudinal size grows parametrically
as $|\Delta x^-|\sim \chi^0\alpha^\prime/0^+$, with $0^+$ the time resolution in the light front coordinate $x^+$~\cite{Susskind:1993aa}
(note that $\Delta x^-\Delta x^+\sim  \alpha^\prime$ by the uncertainty principle).
Parton as wee string bits do not behave as normal matter under Lorentz boost.

\begin{figure}[!htb]
\includegraphics[height=6cm]{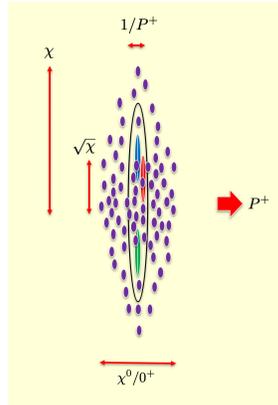}
 \caption{Boosted nucleon with large longitudinal momentum $P^+\sim e^{\frac 12 \chi}$.
 The confined  quark-diquark  pair is highly contracted, with a vanishingly small
 longitudinal size $1/P^+$, and fixed transverse size $\chi^0$. It is surrounded by a halo
 of partons as string bits, which extends transversely as $\sqrt{\chi}$ (diffusive regime) and up to $\chi$ (ballistic regime).
 The halo remains parametrically large longitudinally as $\chi^0/0^+$, with $0^+$ the time resolution along $x^+$~\cite{Susskind:1993aa}. Throughout,
it is described by a continuous NG string with longitudinal momentum $P^+$. All dimensions are in string units.}
  \label{fig:BOOSTEDHAD}
\end{figure}

The  number of wee-string-bits grows  exponentially with
${\bf WW}_{\rm 2PI}\sim e^{\alpha_{\mathbb P}\chi}$.
This  growth is similar to the growth of the longitudinal light front momentum $P^+\sim \gamma\sim e^{\frac 12 \chi}$,
of the boosted hadron. The string growth persists, even though the total cross section saturates by quantum
shadowing, with $\rho(\chi)$ the number of string bits per light front volume $|\Delta x_\perp|| \Delta x^-|$,
\bea
{\bf diffusive\, \,regime:} \,\,b_\perp\sim\sqrt{\alpha^\prime\chi}&\qquad&\rho(\chi)\sim
\frac{e^{\alpha_{\mathbb P}\chi}}{\chi \alpha^{\prime 2}/0^+} \ , \nonumber\\
{\bf  ballistic\, \,regime:}\,\, b_\perp\sim \sqrt{\alpha^\prime}\chi&\qquad& \rho(\chi)\sim
\frac 1{\chi^2\alpha^{\prime 2}/0^+} \ .
\eea
An illustration of this spatial growth under boosting of the nucleon is shown in Fig.~\ref{fig:BOOSTEDHAD}.
The ballistic regime  dominates the total cross section.

\begin{figure}[!htb]
\includegraphics[height=4cm]{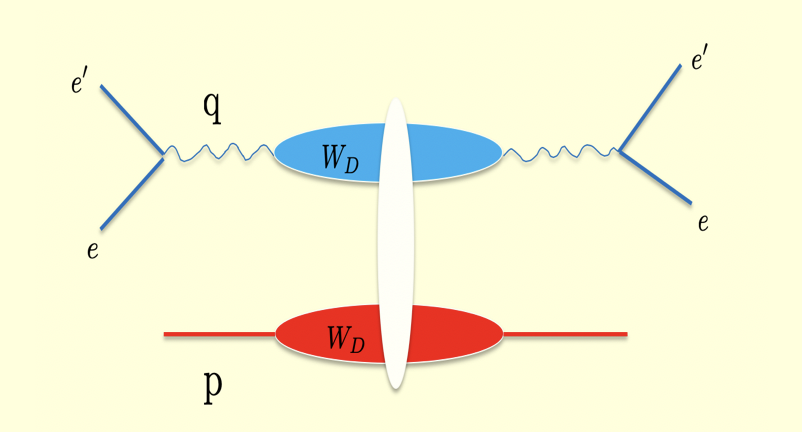}
 \caption{DIS scattering as two Wilson loops ${\bf W}_D$ exchanging a closed NG string,  in the Regge limit.}
  \label{fig:StringDIS}
\end{figure}

These features are accessible to DIS scattering at large $Q^2/m_H^2\gg 1$ and small parton faction $x\ll  1$,
where the virtual photon can be viewed  as a small projectile dipole
of size $a_P\sim 1/\sqrt{Q^2}$, scattering off a target hadron also as a dipole of a larger size $a_T$, as illustrated in Fig.~\ref{fig:StringDIS}.
We recall the  DIS kinematics
$$s-m_H^2={Q^2}\left(\frac 1x -1\right)$$ with the identification $\chi\rightarrow {\rm ln}{\frac 1x}$.
For a fixed target size,
(\ref{ww1X}) translates to
\bea
\label{WWD}
{\bf WW}_{\rm 2PI}(Q^2,x, b)\sim g_s^2 \frac{ 1}{\sqrt{\alpha^\prime Q^2}}
\frac 1{x^{{\alpha_{\mathbb P}(0)}}} e^{-\frac{{\bf b}_\perp^2}{2\alpha^\prime {\rm ln}\frac 1x}}\sim
 \left(\frac{Q^2(x)}{Q^2}\right)^{\frac 12} e^{-\frac{{\bf b}_\perp^2}{2\alpha^\prime {\rm ln}\frac 1x}} \ ,
\eea
after re-insertion of the pre-exponent, with
\bea
Q(x)=\frac{g^2_s}{\sqrt{\alpha^\prime}}\frac 1{x^{{\alpha_{\mathbb P}(0)}}} \ .
\eea
The arguments presented in Appendix~\ref{APP} show that the $F_2$ structure function is
\bea
\label{F2X}
F_2(x, Q^2)\sim xG_{\mathbb P}(x, Q^2)\sim  \left(\frac{Q^2(x)}{Q^2}\right)^{\frac 12} \ ,
\eea
at low-x, where $Q(x)$ maybe regarded as the stringy analogue of the so-called saturation
momentum, with differences with the original proposal by Golec-Biernat-Wusthoff  (GBF)~\cite{Golec-Biernat:1999qor}.
The standard condition for saturation is set by the requirement that ${S}(s,a,b)|_S=e^{-\frac 12}$ in (\ref{SBX})
(a drop by one standard deviation in the GBF Gaussian proposal for $S$), or equivalently
\bea
\label{XSATX}
\frac{d\sigma}{d^2b}\bigg|_S=2(1-e^{-\frac 12})=0.79\qquad\longrightarrow\qquad 14<\chi_S={\rm ln}\frac 1{x_S}<20 \ .
\eea
The rightmost result follows numerically from  the  black-dashed curve in Fig.~\ref{fig:SIGMAABX}, for the impact parameter $b$  in the range $2<b/l_s<6$.
This stringy estimate puts a lower bound on parton-x at saturation $x_S>10^{-6}$, which falls outside the reach of current colliders, including the future EIC.
This conclusion is specific to the Pomeron as the exchange of non-critical NG string in $D=4$ dimensions.

Finally, (\ref{WWD}) may be viewed as the number of wee string bits with parton-x, at a distance $b_\perp=\sqrt {{\bf b}_\perp^2}$ in the transverse
plane, surrounding a fast moving hadron  sourced by a fixed size dipole. The number is small for $Q^2\gg Q^2(x)$, whatever
$b_\perp$. It is large for $Q^2\ll Q^2(x)$, only in  the  disc  ${{b}_\perp}\sim \sqrt{\alpha^\prime{\rm ln}\frac 1x}$, which is seen to
grow diffusively   in the immediate surrounding of the target dipole. It drops substantially in the much wider corona
${{b}_\perp}\sim \sqrt{\alpha^\prime}{\rm ln}\frac 1x$, where the growth is ballistic.

\section{Entanglement in scattering}
\label{SEC_V}
In the large rapidity limit, (\ref{ww1}) is dominated by the tachyon contribution in the closed string exchange
described by the Nambu-Goto string. The tachyon as a mode encodes the quantum entanglement
 between the projectile and the target, carried geometrically by the worldsheet. A way to quantify this, is to recast  (\ref{ww1}) in the form
(\ref{WW3}). This  is readily  identified as the {free energy} of $D_\perp$ massless bosons trapped in a box of size $b$
at temperature $1/\beta$ as illustrated in Fig.~\ref{fig:StringWS}, with~\cite{Liu:2018gae}
\bea
\label{EB1}
S_{\rm 1loop}=\beta F_B=D_\perp\int \frac{b dp}{2\pi}\,{\rm ln}\bigg(1-e^{-\beta |p|}\bigg) \ .
\eea
As a result, the exchange in Fig.~\ref{fig:StringWS} carries a quantum or entanglement {bosonic} entropy
\bea
\label{EB2}
S_{EB}=\beta^2\frac{\partial F_B}{\partial \beta}=D_\perp\int \frac{b dp}{2\pi}\,\frac{2\beta |p|}{e^{\beta |p|}-1}=\frac{D_\perp}6\chi=2\alpha_{\mathbb P}(0)\chi \ .
\eea
Eq. (\ref{EB2}) clearly captures the entropy of fluctuating string bits (gluonic dipoles) on the instanton worldsheet of size $\beta\times b$.

\subsection{Fermionic contribution to the Pomeron intercept and DIS}

\begin{figure}[!htb]
\includegraphics[height=4cm]{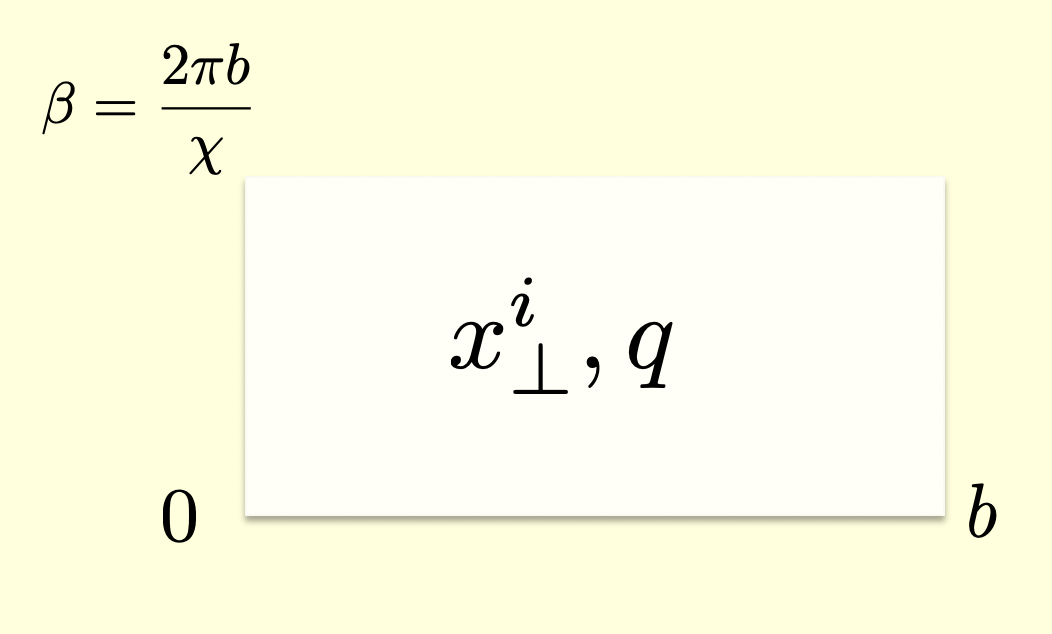}
 \caption{String worldsheet exchange $\beta\times b$ in the Regge limit. The  transverse
fluctuations $x_\perp^i$ with $i=1,..,D_\perp$, and the $n_f$ massless fermions  $q$, are subject to periodic
boundary conditions in $\beta$.}
  \label{fig:StringWS}
\end{figure}

The fermionic correction to (\ref{EB2}) follows immediately from this physical observation, as $n_f$ massless worldsheet fermions also trapped
in $\beta\times b$ as illustrated in Fig.~\ref{fig:StringWS}, with the result
\bea
\label{EB3}
\beta F_F=n_f\int \frac{b dp}{2\pi}\,{\rm ln
}\bigg(1+e^{-\beta |p|}\bigg) \ ,
\eea
in total analogy with (\ref{EB1}). The corresponding quantum or entanglement {fermionic} entropy is then
\bea
\label{EB4}
S_{EF}=\beta^2\frac{\partial F_F}{\partial \beta}=n_f\int \frac{b dp}{2\pi}\,\frac{2\beta |p|}{e^{\beta |p|}+1}=\frac 12 \frac{n_f}6\chi \ ,
\eea
with a net entanglement entropy
\bea
\label{EB4X}
S_{EE}=S_{EB}+S_{EF}=\bigg(1+\frac 12 \frac{n_f}{{D_\perp}}\bigg) \frac{D_\perp}{6}\chi \ ,
\eea

The result (\ref{EB4X}) implies that the stringy Pomeron intercept is affected by the fermionic
corrections on the worldsheet. More specifically, the Pomeron contribution in hadron-hadron
scattering is modified, with  a shifted intercept
\bea
\label{EB5}
\alpha_{\mathbb P}(t)=
 \frac{D_\perp}{12}+\frac{\alpha^\prime}2 t
\rightarrow \tilde{\alpha}_{\mathbb P}(t)=\bigg(1+\frac 12 \frac{n_f}{{D_\perp}}\bigg) \frac{D_\perp}{12}+\frac{\alpha^\prime}2 t \ .
\eea
due to the fermionic contribution.

Also in DIS, we can re-interpret the gluonic $F_2$ structure function (\ref{F2X}) at low-x as
\bea
\label{TF2X}
F_2(x, Q^2)\sim xG_{\mathbb P}(x, Q^2)\sim \frac{1}{x^{\tilde\alpha_{\mathbb P(0)}}} \ ,
\eea
at the resolution fixed by the probing dipole size $Q\sim 1/a$. Note that (\ref{EB4X}) is still solely given by
the gluon density (\ref{TF2X})  at low-x
\bea
\label{EEDIS}
S_{EE}(x,Q^2)\sim {\rm ln}(xG_{\mathbb P}(x, Q^2) \ ,
\eea
albeit with a fermion corrected gluonic intercept. At low-x, the partonic evolution
does not follow from DGLAP, but rather BFKL (weak coupling) or surfaces (strong coupling).
An alternative proposal to account for the fermionic contribution to the entanglement entropy
at low-x was suggested in~\cite{Kharzeev:2021yyf,Hentschinski:2021aux,Hentschinski:2022rsa}.


\begin{figure}[!htb]
\includegraphics[height=6cm]{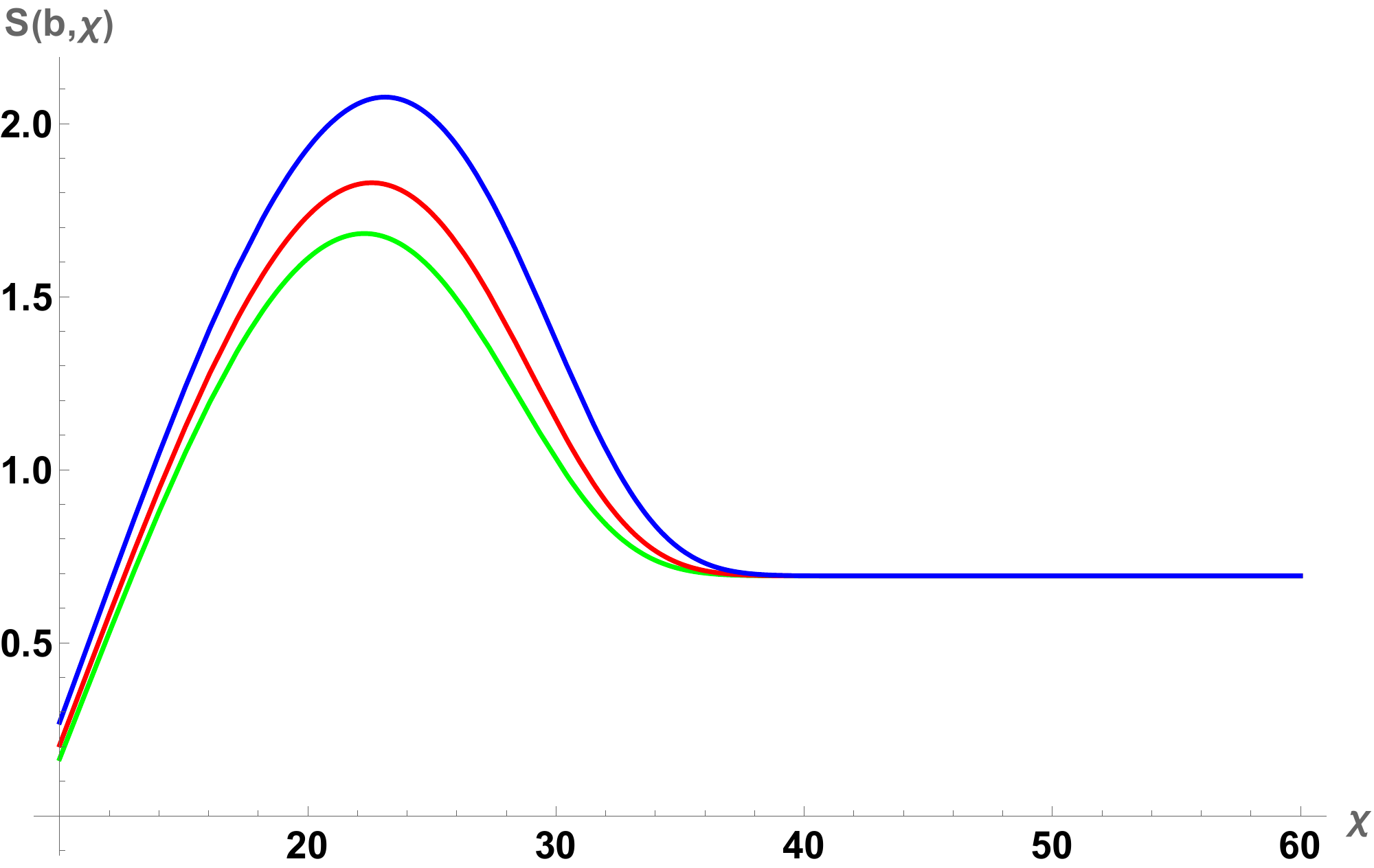}
 \caption{The entanglement entropy between two light-light scattering dipoles in the Regge limit,
 versus the rapidity $\chi$, following from the 2PI NG string contribution to the total cross section shown in Fig.~\ref{fig:SIGMAAB}.
The  green-solid lower  curve, red-solid middle curve and  blue-solid upper are for impact parameters $b=2,4,6$,
for a pair of dipoles of fixed size $a=1$,  in units of the string length. The rapid and linear rise in the entanglement entropy
with rapidity, is stopped and reversed by quantum shadowing. It levels off asymptotically when the Froissart bound is reached.}
  \label{fig:SIGMAAB}
\end{figure}



\subsection{Entanglement and Froissart bound}

Finally, we suggest that in Reggeized hadron-hadron scattering at the Froissart bound, the entanglement
entropy saturates by quantum shadowing, even though the entanglement entropy as measured by (\ref{EEDIS})
in DIS does not.  For that, we interpret the 2PI stringy exchanges in the shadowing process in (\ref{HAD3}), as a
net quantum free energy
\bea
\label{FPROPOSAL}
F_{\rm 2PI}=-\frac 1\beta\,{\rm ln}\bigg(2\big(1-e^{{\bf WW}_{\rm 2PI}}\big)\bigg) \ .
\eea
Note that it reduces to the stringy free energy for small ${\bf WW}_{\rm 2PI}$ and large
rapidity.  Hence the 2PI quantum or entanglement  entropy
\bea
\label{SAT1}
S_{\rm 2PI}=\beta^2\frac{\partial F_{\rm 2PI}}{\partial \beta}={\rm ln}\big(1-e^{{\bf WW}_{\rm 2PI}}\big)+\beta \frac{\partial_\beta{\bf WW}_{\rm 2PI}}{1-e^{-{\bf WW}_{\rm 2PI}}} +{\rm ln}2\rightarrow{\rm ln}2 \ ,
\eea
which is seen to asymptote a constant for fixed $b$ and large rapidity $\chi\gg 1$. In the unitarity limit,
the entanglement is that of a single qubit! Recall that in the black disc limit, the scattering choice appears to be binary, as the elastic and inelastic
cross sections are equal to the classical cross section (Babinet theorem).

In Fig.~\ref{fig:SIGMAAB} we show the entanglement entropy versus rapidity,  using our proposal (\ref{FPROPOSAL}) for the
2PI contribution,  for different values of the impact parameter $b$.
The  green-solid lower  curve, red-solid middle curve and  blue-solid upper curve, are for impact parameters $b=2,4,6$
and fixed dipole size $a=1$, all units of the string length. We have fixed the string coupling $g_s=1$. The rapid initial
rise with rapidity, is caused by the NG tachyon in the single string exchange (\ref{EB4X}). This rise overshoots the unitarity
line, before it is overtaken by quantum shadowing in (\ref{SAT1}), to level off at the Froissart bound.
This levelling off is generic of chaotic systems in their approach to equilibrium~\cite{Latora_1999,Liu:2018gae},
although here from above and not below.

\section{NG tachyon diffusion in a confining warped space}
\label{SEC_IV}
The scattering amplitude of two fixed dipoles of size $a$ in  (\ref{scatteringamplitudeX}),
is dominated by the exchange of the NG tachyon
\bea
\label{TACH}
\exp \left(-\chi\left[\frac{\alpha^\prime}2\bigg( {\bf q}_\perp^2+M_0^2\bigg)\right]\right) \ ,
\eea
in the transverse  $D_\perp$-dimensions,  where the rapidity $\chi$ emerges as a  {\it proper time}.
This exchange is diffusive, and can be  recast (\ref{TACH}) using the contour integral
\bea
\label{TACHCONT}
\int_{-i\infty}^{+i\infty}\frac{dj}{2i\pi}\frac{e^{\chi j}}{j+\frac{\alpha^\prime}2 (M_0^2+{\bf q}_\perp^2)}
=\frac 2{\alpha^\prime}\int_{-i\infty}^{+i\infty}\frac{dj}{2i\pi}{e^{\chi j}} G(j, {\bf q}_\perp) \ .
\eea
The propagator in the complex j-plane satisfies
\bea
\label{GQ}
({\bf q}_\perp^2+m_j^2)\,G(j,{\bf q}_\perp)=1 \ .
\eea
with the j-dependent mass $m_j^2=\frac 2{\alpha^\prime}(j-j_0)$ and $j_0=\frac {D_\perp}{12}$.
This stringy result captures the exchange of an emergent spin-j in the Regge limit,  for  fixed momentum
transfer ${\bf q}_\perp$,
sourced by a projectile and a target dipole of fixed sizes (both set to $a$).

\subsection{Warped  diffusion}
In QCD a boosted hadron is a collection of wee dipole of various sizes as we discussed earlier. The scattering between
these dipoles is mediated by a non-critical NG string. Large size wee dipoles scattering mostly in the IR, through the exchange
of a closed NG string in the confining regime. Small size dipoles scatter mostly in the conformal regime, albeit at strong coupling
in our case. The chief results are different transverse growth size of the string,  at the origin of the transverse size
dependence of the saturation scale: 1/ in the confining regime the growth is linear in the root of the rapidity as in (\ref{WWD});
2/ in the conformal regime, the growth is exponential in the rapidity as we show below.

With this in mind,   the sourcing dipole sizes vary as well, with the combined  change in $a, b_\perp$, expected
to be conformal in the UV, and stringy in the IR.  To realize this, we rewrite (\ref{GQ}) in a general
coordinate space
\begin{align}
\label{DIFX}
-\frac{1}{\sqrt{|g|}}\partial_{\mu}(\sqrt{|g|}g^{\mu\nu}\partial_{\nu}G)+m_j^2G=\frac{1}{\sqrt{|g|}}\delta(z-z')\delta^{D-1}(\vec{x}-\vec{x}') \ ,
\end{align}
by combining $(a, {\bf b}_\perp)\rightarrow (z, \vec{x})=x^\mu$  in $D=1+D_\perp$ space.
But what is the metric $g_{\mu\nu}$ when $z$ is added as a coordinate? (\ref{DIFX}) can be viewed as an evolution equation in our
QCD analysis of the scattering amplitude, with the evolution taking place in $z\sim 1/\sqrt{Q^2}$ and rapidity $\chi$.  Hence,
the metric should exhibit conformal symmetry for small $z$. Inspired by holography, we fix $g_{\mu\nu}$ through the line element
\begin{align}
\label{METRIC}
ds^2=\frac{R^2}{z^2}e^{\mp\kappa^2z^2}(dz^2+d^2{x}_\perp) \ .
\end{align}
For small size dipoles $z\rightarrow 0$, (\ref{METRIC}) reduces to that of AdS$_3$ which is conformal.
For large size dipoles $z\rightarrow 1/\kappa$ as expected from confinement for both  warping  signs.
Although (\ref{METRIC}) is reminiscent of the holographic analysis of the Pomeron in AdS$_5\times$S$_5$~\cite{Polchinski:2002jw,Polchinski:2002jw}, we emphasise that
the present  construction
is not holographic. The starting point is the NG string in flat space with a tachyon for $D_\perp=2$, with no reference to type IIB string theory
in 10-dimensions with no  tachyon~\cite{Ammon:2015wua} (and references therein).  As we noted earlier, the NG string in 4-dimension is the only effective string model currently supported
by QCD lattice simulations.
We define
\begin{align}
\frac{R^2}{\alpha^\prime}\equiv \sqrt{\lambda} \ ,
\end{align}
with $\kappa, R$ to be tied below. Since our approach is not holographic, the identification $\lambda=g_Y^2N_c$ does not follow.
However, it is natural to expect that $R/l_s\gg 1$, since $R$ is the radius of the hyperbolic space, where the warped evolution
of the NG tachyon is justified for large transverse separations.

With this in mind, and using  (\ref{METRIC}) in (\ref{DIFX}), we obtain
\bea
&&-\partial_z^2G(z,z^\prime,t)+(D-2)\bigg(\frac{1}{z}\pm\kappa^2z\bigg)\partial_z G(z,z^\prime,t)\nonumber\\
&&+\bigg(t+\frac{S}{z^2}e^{\mp\kappa^2z^2}\bigg)G(z,z^\prime,t)=z^{D-2}e^{\pm (\frac D2-1)\kappa^2z^2}\,\delta(z-z') \ .
\eea
with $\partial_\perp^2\rightarrow t$ and
\begin{align}
S\equiv S_j=2\sqrt{\lambda}(j-j_0)  \ .
\end{align}
To  remove the first order derivative, we redefine
\begin{align}
\label{REDF}
G(z,z^\prime,t)\rightarrow z^{\frac{D-2}{2}}e^{\pm\frac{D-2}{4}\kappa^2z^2}z^{'\frac{D-2}{2}}e^{\pm\frac{D-2}{4}\kappa^2z^{'2}}G(z,z^\prime,t) \ ,
\end{align}
set  $u=\kappa z$,  expand $e^{\mp u^2}=1\mp u^2+{\cal O}(u^4)$ (moderatly small size dipoles)  to obtain
\bea
&&-\frac{d^2}{du^2}G(u,u^\prime, t)+\nonumber\\
&&\bigg(\frac{S_j+\frac{D(D-2)}{4}}{u^2}+\frac{u^2}{4}(D-2)^2+\frac{t}{\kappa^2}\mp S_j\pm\frac{1}{2}(D-2)(D-3)\bigg) \
G(u,u^\prime, t)=\delta(u-u') \ .\nonumber \\
\eea
For $D=1+D_\perp=3$, we have explicitly
\begin{align}\label{eq:Greenu}
-\frac{d^2}{du^2}G(u,u^\prime, t)+\bigg(\frac{S_j+\frac{3}{4}}{u^2}+\frac{u^2}{4}+\frac{t}{\kappa^2}\mp S_j\bigg)G(u,u^\prime, t)=\delta(u-u') \ .
\end{align}

\subsection{Repulsive warping}
The repulsive warping  with $e^{+\kappa^2z^2}$ acts as absolute confinement for the dipole sizes  in hyperbolic space, characterizing the evolution. (In holography, it is
a regulated hard wall in bulk AdS). In this case the linear and homogeneous equation (\ref{eq:Greenu}) becomes
\begin{align}
\label{GUU}
-\frac{d^2}{du^2}G(u)+\bigg(\frac{S+\frac{D(D-2)}{4}}{u^2}+\frac{u^2}{4}(D-2)^2+\frac{t}{\kappa^2}+S-\frac{1}{2}(D-2)(D-3)\bigg)G(u)=0 \ .
\end{align}
To simplify the equation,  we  consider $u\rightarrow u'=\sqrt{D-2}u=\sqrt{D-2}\kappa z$ , for which (\ref{GUU}) reads
\begin{align}
\label{GWALL}
-\frac{d^2}{du^2}G(u)+\bigg(\frac{S+\frac{D(D-2)}{4}}{u^2}+\frac{u^2}{4}+\tilde t+\tilde S-\frac{1}{2}(D-3)\bigg)G(u)=0 \ .
\end{align}
With
\begin{align}
\label{DEFX}
D_\perp=D-1 \ , \   \tilde t=\frac{t}{(D-2)\kappa^2} \ , \ \tilde S=\frac{S}{D-2} \ .
 \end{align}
the general solutions are of the form
\bea
\label{HYPER}
G_1(u)&=&e^{-\frac{u^2}{4}}u^{1+\sqrt{\frac{D_\perp^2}{4}+(D_\perp-1)\tilde S}}\,{\mathbb M}\left(\frac{2-\frac{D_\perp}{2}+\tilde S+\tilde t+\sqrt{\frac{D_\perp^2}{4}+(D_\perp-1)\tilde S}}{2},1+\sqrt{\frac{D_\perp^2}{4}+(D_\perp-1)\tilde S},\frac{u^2}{2}\right) \ , \nonumber\\
G_2(u)&=&e^{-\frac{u^2}{4}}u^{1+\sqrt{\frac{D_\perp^2}{4}+(D_\perp-1)\tilde S}}\, {\mathbb U}\left(\frac{2-\frac{D_\perp}{2}+S+\tilde t+\sqrt{\frac{D_\perp^2}{4}+(D_\perp-1)\tilde S}}{2},1+\sqrt{\frac{D_\perp^2}{4}+(D_\perp-1)\tilde S},\frac{u^2}{2}\right) \ , \nonumber \\
\eea
where ${\mathbb M}(a,b,z)$ and ${\mathbb U}(a,b,z)$ are the Kummer ${\mathbb M}$ function and the Tricomi ${\mathbb U}$ function, respectively.
${\mathbb M}$ is regular at $u=0$, while ${\mathbb U}$ has a branch cut at $u=0$. The solution to (\ref{eq:Greenu}) reads
\bea
\label{HYPER12}
G(u,u^\prime)&=&{\cal A}\,G_2(u)G_1(u^\prime)\qquad\qquad u>u^\prime \ , \nonumber\\
G(u,u^\prime)&=&{\cal A}\,G_1(u)G_2(u^\prime)\qquad\qquad u<u^\prime \ .
\eea
with ${\cal A}$ given by the Wronskian
\begin{align}
{\cal A}=\frac 1{{{\cal W}(G_1,G_2)}}\sim-{\Gamma\left(\frac{2-\frac{D_\perp}2+\tilde t+\tilde S+\sqrt{(D_\perp-1)\tilde S+\frac{D_\perp^2}{4}}}{2}\right)} \ .
\end{align}

\subsection{Reggeized trajectories}

${\cal A}$ has poles when
\begin{align}
\frac{\tilde t+\tilde S+2-\frac{D_\perp}{2}+\sqrt{(D_\perp-1)\tilde S+\frac{D_\perp^2}{4}}}{2}=-n \ .
\end{align}
or squared masses  $|t(j,n)|$ given by
\bea
|\tilde t(j,n)|=2n+\tilde S+2-\frac{D_\perp}{2}+\sqrt{(D_\perp-1)\tilde S+\frac{D_\perp^2}{4}} \ .
\eea
By extending the tachyon diffusion to AdS$_3$ space, the original tachyon  pole
morphs into a multitude of Regge poles,
\begin{align}
\tilde S=\frac{2\sqrt{\lambda}(j-j_0)}{D_\perp-1}=
\frac{-5+2D_\perp-4n+\frac{2|t|}{(D_\perp-1)\kappa^2}\pm \sqrt{(2D_\perp-3)^2-8(D_\perp-1)n+\frac{4|t|}{\kappa^2}}}{2} \ .
\end{align}
For $n=0$ and small $t$, the Regge trajectories for the $\pm$ signs are
\bea
\label{GOOD}
j_+&=& j_0+\frac{(D_\perp-1)(D_\perp-2)}{\sqrt\lambda}+
\bigg(\frac {3D_\perp-4}{D_\perp-2}\bigg)\frac{\alpha^\prime}2 |t|
+{\cal O}(t^2) \nonumber\\
j_-&=& j_0-\frac{D_\perp-1}{2\sqrt\lambda}+\frac{\alpha^\prime}2 |t| +{\cal O}(t^2) \ .
\eea
provided that $$\kappa^2R^2=\frac{D_\perp-2}{2D_\perp-3}$$ with $\sqrt{\lambda}=R^2/\alpha^\prime$.
The dominant contribution to the Pomeron stems from the $j_-$ trajectory, which is the closest to the origin
in the j-plane. The warping shifts down the flat space intercept $j_0$.

\subsection{Confining regime}

For large $t$ and large $n$ the poles in the j-plane become imaginary, but with negative real parts.
To proceed with the contour integration in (\ref{TACHCONT}) for the tachyon propagator, we select the branch cut of  $\sqrt{S+1}$ from $-1-i\infty$ to $-1$. With this in mind, the dominant contribution stems from the first pole $n=0$
with $j=j_-$.
At the pole, the hypergeometric functions in (\ref{HYPER}) simplify
\bea
\label{SYMP}
&&{\mathbb M}\bigg(0, 1+\sqrt{S+\frac{D_\perp^2}4}, \frac {u^2}2\bigg) = 1\nonumber\\
&&{\mathbb U}\bigg(0, 1+\sqrt{S+\frac{D_\perp^2}4}, \frac {u^2}2\bigg)=1
\eea
Using (\ref{HYPER},\ref{HYPER12},\ref{SYMP}) in (\ref{TACHCONT}) and carrying the j-integral gives  the  warped tachyon propagator
\begin{align}
G(z, z^\prime, b_\perp) \sim (zz^\prime)^{D_\perp -\frac 12}\int \frac{d^2q}{(2\pi)^2} \,
e^{\chi(j_0-\frac{D_\perp -1}{2\sqrt\lambda}-\frac{\alpha' q^2}2)+iq\cdot b}
\end{align}
after re-winding the re-definition (\ref{REDF}),
for large $\chi$. The result for the warped tachyon propagator is
\bea
\label{GZZX}
G(z, z^\prime, b_\perp) \sim (zz^\prime)^{D_\perp -\frac 12}
e^{\chi(j_0-\frac{D_\perp -1}{2\sqrt\lambda})-\frac{b^2_\perp}{2\chi\alpha'}}
\eea
 (\ref{GZZX}) reduces to the unwarped result with a shifted down  intercept.

\subsection{Conformal regime}

The effects of the warping is mostly in action away from the confining regime. Indeed,
in the conformal regime, the hypergeometric functions are limited to small $u=\kappa z$
and large $t\gg \kappa^2$, which we will refer to as the conformal limit. More specifically, this
amounts to the limits
\begin{align}
\lim_{\kappa \rightarrow 0} G_1(u) , \ G_2(u) \ ,
\end{align}
which are not simply the $u\rightarrow 0$ limits, because $\frac{t}{\kappa^2}$ in the second argument has to go to infinity. To obtain these limits, the simplest way is to note that the differential equation (\ref{GWALL}) reduces to
\begin{align}
-\frac{d^2}{dz^2}G(z)+\bigg(\frac{S+\frac{D(D-2)}{4}}{z^2}+t\bigg)G(z)=0 \ ,
\end{align}
with two solutions
\begin{align}
\tilde G_1(z)=\sqrt{z}J_{-\sqrt{S+\frac{D_\perp^2}{4}}}(-i\sqrt{t}z) \ , \\
\tilde G_2(z)=\sqrt{z}Y_{-\sqrt{S+\frac{D_\perp^2}{4}}}(-i\sqrt{t}z) \ .
\end{align}
Here  $-\sqrt{S+\frac{D_\perp^2}{4}}$ is chosen to have a negative real part. With this in mind, the warped tachyon propagator
in the conformal limit is of the form
\begin{align}
G(z, z^\prime, b_\perp)=\frac{\pi t}{2}\sqrt{zz^\prime}J_{-\sqrt{S+\frac{D_\perp^2}{4}}}(-i\sqrt{t}z_{<})Y_{-\sqrt{S+\frac{D_\perp^2}{4}}}(-i\sqrt{t}z_{>}) \ .
\end{align}
For  $b_\perp\gg z,z^\prime$,  the Bessel functions reduce to
\bea
J_{-\sqrt{S+\frac{D_\perp^2}{4}}}(-i\sqrt{t}z)
&\rightarrow& \frac{1}{\Gamma\bigg(1-\sqrt{S+\frac{D_\perp^2}{4}}\bigg)} \bigg(\frac{\sqrt{t}z}{2}\bigg)^{-\sqrt{S+\frac{D_\perp^2}{4}}} \ , \\
Y_{-\sqrt{S+\frac{D_\perp^2}{4}}}(-i\sqrt{t}z_{>})&\rightarrow& \cos \bigg(\pi\sqrt{S+\frac{D_\perp^2}{4}}\bigg)
\Gamma\bigg(\sqrt{S+\frac{D_\perp^2}{4}}\bigg)\bigg(\frac{\sqrt{t}z}{2}\bigg)^{-\sqrt{S+\frac{D_\perp^2}{4}}} \ .
\eea
As a result, the warped tachyon propagator  in the conformal regime is dominated by the following S-integral
\begin{align}
G(z, z^\prime, b_\perp) \sim \int \frac{d^2q}{(2\pi)^2}e^{i{q}\cdot n_\perp}\int dS\exp \bigg(\chi\bigg(j_0+\frac{S}{2\sqrt{\lambda}}\bigg)+\bigg(\frac{1}{2}-\sqrt{\frac{D_\perp^2}{4}+ S}\bigg)
\ln \bigg(\frac{zz^\prime q^2}{2b_\perp^2}\bigg)\bigg)\ ,
\end{align}
with $n_\perp$ a unit vector along $q_\perp$ after rescaling away the transverse momentum,
and trading $j\rightarrow \tilde S$ using (\ref{DEFX}). The result is
\bea
\label{GZZCONF}
G(z, z^\prime, b_\perp)\sim \frac{\xi}{(4\pi{\cal D} \chi)^{\frac{3}{2}}}\exp \bigg(\chi\bigg(j_0-\frac{{\cal D}D_\perp^2}{4}\bigg)-\frac{\xi^2}{4 \chi \cal D}\bigg) \ ,
\eea
with the conformal variable $\xi=\ln \frac{b_\perp^2}{zz^\prime}$ and diffusion like constant ${\cal D}=\frac{1}{2\sqrt{\lambda}}$.

\subsection{Cross section in conformal regime}

In terms of (\ref{GZZCONF}), the dipole-dipole scattering amplitude in the conformal regime is now
\begin{align}
\label{WWCONF}
{\bf WW}_{\rm 2PI}=-\frac{g_s^2}{4}(2\pi)^{\frac{3}{2}}G=-\frac{g_s^2\xi}{4(2{\cal D}\chi)^{\frac{2}{3}}}e^{\tilde j_0\chi-\frac{\xi^2}{4\chi{\cal D}}} \ ,
\end{align}
with $\tilde j_0=j_0-\frac 14 {\cal D}D_\perp^2$. Inserting (\ref{WWCONF}) in the 2PI contribution to the total cross section (\ref{HAD3}) gives
\begin{align}
\label{SIGCONF}
\sigma(s)=2\int d^2b\bigg(1-e^{{\bf WW}_{\rm 2PI}}\bigg)=2\pi zz' \int_{0}^{\infty} d\xi e^{\xi} \bigg(1-\exp \left(-\frac{g_s^2\xi}{4(2{\cal D}\chi)^{\frac{2}{3}}}e^{\tilde j_0\chi-\frac{\xi^2}{4\chi{\cal D}}} \right)\bigg) \ .
\end{align}
For  $\tilde j_0<4{\cal D}$, the integral in (\ref{SIGCONF}) can be undone for large $\chi$, with the result for the total cross section
in the conformal regime
\begin{align}
\label{SSXX}
\sigma(s) \rightarrow 2\pi zz' \frac{g_s^2\sqrt{\pi}}{2\sqrt{2}}e^{\chi\bigg(j_0-{\cal D} \bigg(\frac{D_\perp^2}4-1\bigg)\bigg)} \ .
\end{align}
It grows exponentially, i.e. $\sigma(s)\sim zz^\prime s^{j_0}$ with $j_0=\frac{D_\perp}{12}=\frac 16$ for $D_\perp=2$, much like the BFKL result in pQCD,
which is also conformal at weak coupling.

Modulo the NG string assignments  for $j_0$ and ${\cal D}$, the contribution (\ref{GZZCONF})  to the Reggeized scattering amplitude,
and the  total dipole-dipole cross section, is analogous to the result following from Mueller$^\prime$s dipole wavefunction evolution
in pQCD~\cite{Mueller:1994gb}. More importantly, the  results (\ref{GZZX}) (confining regime) and (\ref{GZZCONF}) (conformal regime),
show that the general and warped NG result (\ref{HYPER}) interpolates continuously between these two regimes in Reggeized scattering.
This point was originally made in the context of the gravity dual construction~\cite{Polchinski:2002jw,Brower:2006ea}, with no tachyon in bulk.

\section{Conclusions}
\label{SEC_VI}
One of the most striking features of the detailed lattice studies of the  fuzzy QCD string,
is its description as a fundamental  NG string for large and even  relatively small lengths. This
observation has been numerically checked in both 3- and 4-dimensions, and for different  SU(N$_c$) realizations.
We have used this observation, to analyze the stringy potential between a pair of static and scattering dipoles, as
well as their quantum entanglement.

The derivation of the static potential of two fixed size dipoles, follows from the exchange of closed strings or
glueballs in the quenched approximation (large $N_c$ limit). Using the NG string, we have shown that this
exchange is dominated by the tachyonic mode and attractive at large distances. The attraction persists at
short distances, albeit in a power law form following from the resummation over the string states. This change
in the static potential is captured by an emergent quantum entropy.

The derivation of the scattering amplitude for two fixed size dipoles can be obtained using similar arguments,
by noting that it follows from a potential between two dipoles set at an angle $\theta$, which is then analytically
continued to the rapidity $i\chi$. In the Regge limit, the scattering amplitude is totally fixed by the tachyonic
mode of the NG string, as all the excited modes are suppressed at large rapidity. There is a total  parallel between
the potential channel and the scattering channel, where the 2PI contribution is retained in both, in leading order
in $1/N_c$. This contribution yields unitarization by saturation of the Froissart bound.

We have extended some results regarding the quantum entanglement as captured
by the NG tachyon in dipole-dipole scattering.  The inclusion of the worldsheet fermions,
modify the intercept of the stringy Pomeron, thereby changing the quantum entanglement as measured
by DIS through solely  a modification of the gluonic density. In the presence of shadowing, the quantum
entanglement entropy is depleted, and saturates at the Froissart bound.

The  NG tachyon contribution to the Reggeized scattering amplitude of two dipoles, captures two key aspects
of the exchanged string: 1/ an exponential growth in the number of string bits at the origin of the growth of the
total cross section at large rapidity prior to saturation; 2/ a diffusive spread of the string bits in the transverse
plane. It is the balance between these two phenomena that yields saturation by quantum shadowing, as captured
by the 2PI contribution.

Perturbative QCD arguments using BFKL evolution of gluons as dipoles, have shown that the gluon sizes evolve
and that the evolution is conformal in the UV. This aspect of QCD in the Regge limit, can be extended to the
exchanged NG string, by considering the diffusive spreading of the string bits in the transverse plane together
with the changes in the source and target dipole sizes. In other words, the NG tachyonic mode should diffuse
in conformal 3-dimensional space (transverse space plus dipole size) as opposed to simply 2-dimensional
space (transverse space).

We have shown how to explicitly extend the diffusion of the NG tachyon mode in flat transverse space,
to curved  AdS$_3$ plus a repulsive wall. We have emphasised that this approach is not holographic,
since no string-gauge duality is used. In the conformal limit, the modified  NG tachyon diffusion in
proper time and curved space, yields results for the scattering amplitude of two dipoles with evolving sizes,
similar to those following from the BFKL evolution of Mueller$^\prime$s wavefunction in QCD.
In the confining regime, the NG  tachyon diffusion is preserved, albeit with a shifted down intercept.

\vskip 1cm
{\bf Acknowledgements}

We thank Krzysztof Kutak  for discussions.
This work is supported by the Office of Science, U.S. Department of Energy under Contract No. DE-FG-88ER40388 and by the Priority
Research Areas SciMat and DigiWorld under program Excellence Initiative - Research University at the Jagiellonian University in Kraków.

\appendix

\section{Relation to pQCD dipoles}
\label{APP}
There are few parallels between the stringy results we have developed, and those established using pQCD.
In particular, the 2PI string amplitude (\ref{HAD3}) at fixed $b$, can be recast in the form
\bea
\label{HAD8}
1-e^{{\bf WW}(s, a, b)}=1-{\rm exp}\bigg(-\frac{g_s^2}4 \frac{a^2}{\alpha^\prime}\,xG_{\mathbb P}(x, 1/a^2) S(b)\bigg) \ ,
\eea
with $g_s=f(\lambda)/N_c$, the Gaussian profile
\bea
\label{HAD9}
S(b)=\bigg(\frac{\pi}\chi\bigg)^{\frac{D_\perp}2}\,e^{-S_{\rm min}}=\bigg(\frac{\pi}\chi\bigg)^{\frac{D_\perp}2}\,e^{-\frac{b^2}{2\chi\alpha^\prime}} \ ,
\eea
and $\chi={\rm ln}\frac 1x$  in DIS. The diffusion of the string bits in the transverse plane is manifest
in (\ref{HAD9}).
(\ref{HAD8}) is very similar to
the Glauber-Mueller formula for multiple dipole-target interactions~\cite{Mueller:1989st}
\bea
\label{HAD10}
N_{\rm GM}(x_{01}, x, b)=1-{\rm exp}\bigg(-\frac{\lambda}{N^2_c} \frac{x_{01}^2}{8R^2}\,xG_{\rm DGLAP}(x, 4/{x_{01}^2}) S(b)\bigg) \ ,
\eea
where the dipole size is $x_{01}$ ($a$ in our case),
$R$ is the radius of the target ($\# \sqrt{\alpha^\prime}$ in our case), and $S(b)$ is identified with the dipole profile function inside the target
((\ref{HAD9}) in our case). (\ref{HAD10}) is an extension of the original Golec-Biernat-Wusthoff formula for the analysis of saturation in DIS at HERA~\cite{Golec-Biernat:1999qor},
through the addition of the profile function.

In general, the forward amplitude ${\cal T}(b_\perp,\chi)$ of dipole-nucleus scattering at impact parameter $b_\perp$ and rapidity $\chi$,
relates to the full S-matrix  ${\cal S}(b_\perp,\chi)$ as
\begin{align}
1-{\cal S}({b}_\perp,\chi)=-i{\cal T}({b}_\perp,\chi) \ ,
\end{align}
and the total cross section is given by
\begin{align}
\sigma (\chi)=2\int d^2b {\rm Re}\left(1-{\cal S}({b}_\perp,\chi)\right) \ .
\end{align}
In our case or in the color glass condensate  model (CGC), the full S-matrix is approximated by the Wilson-loop average
\begin{align}
{\cal S}=\frac{1}{N_c}\langle {\rm Tr}V_{\vec{x}_0}V^{\dagger}_{\vec{x}_1} \rangle_{\rm target} \ ,
\end{align}
with $V_{\vec x}$ a Wilson line located at $\vec x$.
If the target is another Wilson-loop, it simply reduces to the ${\bf WW}$ corrector. In the CGC model, this Wilson-loop average exponentiates~\cite{Kovchegov:2012mbw}
(and references therein)
\begin{align}
{\cal S}=\exp \bigg(-\frac{1}{4}{x}^2_\perp Q_s^2({b}_\perp,Y)\ln \frac{1}{|{x}_\perp|\Lambda}\bigg)  \ .
\end{align}
Our large ${b}_\perp$ result cuts off the growth by a delicate balance
 between the factors $e^{\frac{D_\perp\chi}{12}}$ (growth of the string bits) and $e^{-\frac{b^2}{2\chi\alpha^\prime}}$ (diffusion penalty).

\bibliography{ENT}

\end{document}